\documentclass{article}
\usepackage[T1]{fontenc}
\usepackage[utf8]{inputenc}
\usepackage{author_kit_ismir/ismir} % Remove the "submission" option for camera-ready version
\usepackage{amsmath,amssymb,author_kit_ismir/cite,url}
\usepackage{graphicx}
\graphicspath{{author_kit_ismir/}}
\usepackage{tabularx}
\usepackage{booktabs}
\newcolumntype{Y}{>{\centering\arraybackslash}X}
\newcolumntype{P}[1]{>{\centering\arraybackslash}p{#1}}
\makeatletter\@ifundefined{KV@Gin@alt}{\define@key{Gin}{alt}[]{}}{}\makeatother
\usepackage{color}
\usepackage{xcolor}
\usepackage{tikz}
\usetikzlibrary{arrows.meta,positioning,fit,calc,shapes.geometric,bending,decorations.pathmorphing}

\definecolor{latfill}{HTML}{DCE6FC}
\definecolor{latline}{HTML}{2B4785}
\definecolor{tokfill}{HTML}{FFE8D4}
\definecolor{tokline}{HTML}{A3541E}
\definecolor{strokegray}{HTML}{2A2A2A}
\definecolor{modelfill}{HTML}{DDD4F0}
\definecolor{modelline}{HTML}{5A4AA0}
\definecolor{dacfill}{HTML}{ECECEC}
\definecolor{dacline}{HTML}{6B6B6B}
\definecolor{TokFill}{HTML}{FFE8D4}
\definecolor{TokEdge}{HTML}{A3541E}
\definecolor{LatFill}{HTML}{DCE6FC}
\definecolor{LatEdge}{HTML}{2B4785}
\definecolor{EncFill}{HTML}{DDD4F0}
\definecolor{EncEdge}{HTML}{5A4AA0}
\definecolor{AggFill}{HTML}{D0EAE5}
\definecolor{AggEdge}{HTML}{2D7F86}
\definecolor{ArrCol}{HTML}{3F5670}
\definecolor{FanCol}{HTML}{A9B4C4}

\tikzset{
  tok/.style   = {draw=tokline, fill=tokfill, text=tokline, rounded corners=1.5pt,
                  minimum size=5.5mm, inner sep=0pt, font=\scriptsize, line width=0.5pt},
  tokd/.style  = {minimum size=5.5mm, inner sep=0pt, font=\scriptsize},
  lat/.style   = {draw=latline, fill=latfill, text=latline, rounded corners=1.5pt,
                  minimum width=4.5mm, minimum height=9mm, inner sep=0pt,
                  font=\scriptsize, line width=0.5pt},
  latd/.style  = {minimum width=4.5mm, minimum height=9mm, inner sep=0pt, font=\scriptsize},
  model/.style = {draw=modelline, rounded corners=2.5pt, text=modelline,
                  minimum width=22mm, minimum height=14mm, align=center, fill=modelfill, line width=0.6pt},
  dacbig/.style  = {draw=strokegray, trapezium, trapezium stretches=true,
                    trapezium left angle=72, trapezium right angle=72, shape border rotate=180,
                    minimum width=30mm, minimum height=9mm, inner sep=2pt, fill=white, line width=0.6pt},
  dacflip/.style = {draw=strokegray, trapezium, trapezium stretches=true,
                    trapezium left angle=72, trapezium right angle=72,
                    minimum width=30mm, minimum height=9mm, inner sep=2pt, fill=white, line width=0.6pt},
  dacm/.style    = {draw=dacline, trapezium, trapezium stretches=true,
                    trapezium left angle=70, trapezium right angle=70,
                    minimum width=22mm, minimum height=10mm, inner sep=2pt, fill=dacfill,
                    text=dacline, font=\small, line width=0.6pt},
  ar/.style      = {-{Latex[length=2.2mm, width=1.8mm]}, line width=0.6pt, color=strokegray,
                    shorten >=1pt, shorten <=1pt, line cap=round},
  arsnake/.style = {thick, decorate, color=strokegray,
                    decoration={snake, amplitude=1.1mm, segment length=3mm}},
}

\title{Geometric Iterative Retrieval for\\ Neural Audio Codec Resynthesis}

\multauthor
{Leo Schmidt-Traub \quad Frédéric Berdoz \quad Luca A. Lanzend\"orfer \quad Roger Wattenhofer} { 
ETH Zurich\\
{\tt\small \{leoschmidt, fberdoz, lanzendoerfer, wattenhofer\}@ethz.ch}
}

\def\authorname{L. Schmidt-Traub, F. Berdoz, L. A. Lanzend\"orfer, and R. Wattenhofer}

\usepackage[bookmarks=false,pdfauthor={\authorname},pdfsubject={\pdfsubject},hidelinks]{hyperref}
\begin{document}

\maketitle

% FINAL
\begin{abstract}
	Neural audio codecs based on Residual Vector Quantization (RVQ) have become the dominant discrete representation for token-based general audio generation, yet resynthesizing high-quality audio from coarse codec tokens remains an open problem and bounds the fidelity of every system that generates them. Prior work has framed resynthesis as a choice between discrete token prediction and continuous regression. We argue that this dichotomy is incomplete and introduce \emph{geometric iterative retrieval}, a paradigm that uses the RVQ layer hierarchy itself as a natural iterative decomposition in continuous codebook space. Rather than classifying over discrete vocabularies or regressing to a single target vector, our method performs contrastive retrieval in the codebook's geometric space. We evaluate our method on codec restoration tasks across speech and music, and show improvements over both single-pass token prediction and one-step regression baselines.
\end{abstract}

\section{Introduction}\label{sec:introduction}

% FINAL
Neural audio codecs based on Residual Vector Quantization~\cite{juang1982rvq,zeghidour2021soundstream,defossez2022encodec,kumar2023dac} have become a key component of modern audio generation. Systems for speech synthesis~\cite{wang2023valle}, music generation~\cite{copet2023musicgen}, and general audio modeling~\cite{borsos2022audiolm} all transform audio into sequences of discrete tokens via RVQ, then generate these tokens with language-model-style architectures. RVQ encodes information at decreasing granularity, where the first codebook captures coarse structure and subsequent layers add progressively finer detail. Most work tackles the problem of RVQ generation by having a large model generate the first layer, and a separate, smaller model generate higher layers from the first.

% FINAL
Liu et al.~\cite{liu2024codec_resyn} showed that the choice of resynthesis strategy significantly impacts output quality. They identified two existing paradigms and their characteristic failure modes: \emph{Token prediction} treats resynthesis as iterative classification over discrete codebook indices, predicting tokens layer by layer. It naturally respects the codec hierarchy but is geometry-blind, treating all incorrect codebook entries as equally wrong regardless of their distance from the target. \emph{One-step regression} operates in continuous space, predicting the full latent representation in a single forward pass. It is geometrically informed but commits to a single global prediction with no iterative correction. Using mean-squared error as an objective for regression also leads to mean-seeking behavior and over-smoothing, which, due to the highly non-linear nature of the neural audio codec's output projection, leads to suboptimal performance. Liu et al. proposed Schr\"{o}dinger Bridge diffusion as a possible solution. However, diffusion imposes an iterative structure (a noise schedule) that is independent of the codec's own hierarchy.

% FINAL
\begin{table}[t]
	\centering
	\small
	\begin{tabularx}{\linewidth}{p{1.6cm} P{1.7cm} Y}
		\toprule
		                    & \textbf{Single-step} & \textbf{Iterative}        \\
		\midrule
		\textbf{Discrete}   & (degenerate)         & Token prediction          \\
		\textbf{Continuous} & Regression           & Diffusion / \textbf{Retrieval (ours)} \\
		\bottomrule
	\end{tabularx}
	\caption{Codec resynthesis methods organized along two axes, the
		prediction space (discrete vs.\ continuous) and the refinement strategy
		(single-step vs.\ iterative). Prior work occupies three of the four
		cells. Our method shares the continuous-iterative cell with diffusion,
		but iterates along the codec's own RVQ hierarchy rather than along an
		externally imposed noise schedule, giving each step a meaning grounded
		in the codec rather than in the sampler.}
	\label{tab:paradigm_grid}
\end{table}

% FINAL
We observe that the design space is not a spectrum from token prediction to regression but a two-dimensional grid (Table~\ref{tab:paradigm_grid}), spanning two axes: whether the prediction space is discrete or continuous, and whether refinement is single-step or iterative. The bottom-right cell, iterative prediction in continuous space, is what we explore in this paper. Diffusion methods occupy this cell with a noise-schedule iteration that is independent of the codec structure. We show that the RVQ hierarchy itself provides a natural, semantically meaningful decomposition that offers an alternative to diffusion's denoising schedule, where each step corresponds to one level of refinement aligned with the codec's own structure.

% FINAL
We introduce \emph{geometric iterative retrieval},\footnote{\url{https://github.com/ETH-DISCO/codec-resynthesis}} a resynthesis paradigm that is distinct from both token prediction and regression. Each of our $D{-}1$ steps predicts a specific RVQ layer with a specific role in the codec hierarchy, trained with a CLIP-style contrastive loss~\cite{radford2021clip} that respects codebook geometry. A self-attention aggregator further replaces the additive assumption of standard RVQ decoding~\cite{borsos2023soundstorm} with learned, non-additive combinations of previously predicted layers. 

% TO BE IMPROVED
\smallskip \noindent Our contributions can be summarized as follows:
\begin{itemize}
	\item We propose geometric iterative retrieval, a method
	      for predicting in the space of per-layer continuous codebooks,
	      trained with a CLIP-style contrastive objective, and
	      conditioned on previous layers via a learned non-additive
	      aggregator that replaces RVQ's additive residual assumption.

	% \item A two-axis framing of the resynthesis design space
	%       (Table~\ref{tab:paradigm_grid}) that separates prediction
	%       space (discrete vs.\ continuous) from refinement strategy
	%       (single-step vs.\ iterative), and identifies
	%       continuous-iterative prediction aligned with the codec's own
	%       RVQ hierarchy as a novel approach.

	\item An empirical study on DAC codec restoration across
	      speech and music. Our method attains the best LSD against
	      every learned baseline and is preferred by listeners in a
	      double-blind human evaluation test over single-pass token-prediction,
	      MSE-regression, and cosine-regression baselines.
\end{itemize}

\section{Related Work}
\label{sec:related_work}

% DONE
Neural audio codecs have become the dominant discrete representation for audio generation and processing, enabling systems that cast speech synthesis, music generation,
and audio enhancement as language modeling over codec tokens~\cite{borsos2022audiolm, wang2023valle, copet2023musicgen}.
These systems typically decompose audio into multiple layers of tokens via Residual Vector Quantization (RVQ),
and then focus their modeling effort on producing the first layer, leaving the recovery of remaining layers as a secondary problem.
Yet the choice of resynthesis strategy has significant impact on output quality~\cite{liu2024codec_resyn}.

\subsection{Neural Audio Codecs and RVQ}
\label{sec:codecs}

RVQ~\cite{juang1982rvq} decomposes a continuous latent $\mathbf{z} \in \mathbb{R}^d$ into $D$ layers of codebook vectors $\mathbf{e}_k \in \mathcal{C}_k = \{\mathbf{c}_k^{(1)}, \dots, \mathbf{c}_k^{(V)}\}$, where $\hat{\mathbf{z}} = \sum_{k=1}^{D} \mathbf{e}_k$;
layer $1$ captures coarse spectral structure and subsequent layers encode progressively finer residual detail.
This paradigm was established by SoundStream~\cite{zeghidour2021soundstream} and extended by EnCodec~\cite{defossez2022encodec}.
The Descript Audio Codec (DAC)~\cite{kumar2023dac}, which we use throughout this work, performs lookup in a low-dimensional space and achieves state-of-the-art fidelity, particularly for music.

% DONE
Liu et al.~\cite{liu2024codec_resyn} showed that one-shot regression on the latent $\mathbf{z}$ performs better than single-pass discrete token prediction.
Our method builds on this insight but takes a different approach:
rather than targeting $\mathbf{z}$ directly, we predict individual codebook vectors $\mathbf{e}_k$ layer by layer.
We use a contrastive learning objective, inspired by CLIP~\cite{radford2021clip}, which showed that contrastive learning over a shared embedding space can replace explicit classification.
We apply the same principle to codebook lookup, treating each prediction as retrieval over codebook entries rather than classification among them.

% UPDATE
Masked generative models such as SoundStorm~\cite{borsos2023soundstorm} and
MaskGCT~\cite{wang2024maskgct} refine tokens over multiple masked-decoding
passes, each refinement remaining a prediction over discrete codebooks. They therefore occupy the same discrete-iterative cell of Table~\ref{tab:paradigm_grid} as our token-prediction, which is why we do not compare against them.

\subsection{Coarse-to-Fine Token Prediction}
\label{sec:token_prediction}

The dominant resynthesis approach treats the problem as iterative discrete classification.
Given tokens from layers $1, \dots, k{-}1$, a model predicts the codebook index for layer $k$ by classifying over $V$ entries using cross-entropy loss.
This paradigm is used in AudioLM's fine acoustic stage~\cite{borsos2022audiolm}, VALL-E's non-autoregressive model~\cite{wang2023valle}, and SoundStorm's parallel decoding~\cite{borsos2023soundstorm}.

% DONE
Token prediction naturally respects the codec hierarchy and is iterative by design.
However, the cross-entropy objective is geometry-blind, treating all incorrect codebook entries as equally wrong.
The objective carries no explicit geometric signal: predicting an entry geometrically close to the target incurs the same loss as predicting one on the opposite side of the codebook space.
% This provides no differentiated gradient signal for near-misses, as errors at each layer are hard: a wrong codebook entry is a wrong vector entirely, 
% with no information about \emph{how} wrong the prediction was for downstream layers to compensate.
For the task of codec resynthesis, a single misprediction in earlier layers propagates down to others, leading to poor reconstruction.
The discrete, single-step variant, predicting all layers independently in parallel, sacrifices inter-layer conditioning and is generally inferior,
due to an exploding vocabulary size.

% \subsection{One-Step Regression (Continuous, Single-Step)}
\label{sec:regression}

% DONE
The alternative operates in continuous space, predicting the full latent representation in a single forward pass using MSE or cosine similarity loss.
Liu et al.
~\cite{liu2024codec_resyn} showed that directly targeting the latent $\mathbf{z}$ via
one-shot regression outperforms coarse-to-fine prediction on codec resynthesis, and Kammoun et al.
~\cite{kammoun2024codec_enhancement} independently confirmed that continuous latent prediction consistently outperforms single-pass discrete token prediction for speech enhancement in the codec space.
Du et al.~\cite{du2023lauragpt} similarly motivate their one-step regression vocoder by the ``multi-modal distribution nature of codec tokens,'' which they identify as the obstacle to iterative discrete prediction.

% DONE
The fundamental limitation of one-step regression is statistical: it encourages mean-seeking behavior rather than committing to a single mode.
Given only coarse tokens, the posterior over fine-grained details is typically multimodal, with multiple plausible high-frequency realizations consistent with the same coarse structure.
The MSE-optimal prediction under a multimodal posterior is the conditional mean, which averages over modes rather than committing to any one of them.
An alternative is predicting the layers of all codebooks in parallel. Since DAC, and neural audio codecs at large, use codebooks with a fixed length at each layer~\cite{kumar2023dac}, we compute the cosine similarity at each layer and quantize based on orientation rather than magnitude.
This approach works significantly better, but has the disadvantage of sacrificing inter-layer conditioning.

\subsection{Diffusion and Flow-Based Methods (Continuous, Iterative, Codec-Agnostic)}
\label{sec:diffusion}
% DONE
Diffusion and flow-based methods work by iterating in the continuous space.
Liu et al.~\cite{liu2024codec_resyn} proposed Schr\"{o}dinger Bridges that transport from a degraded codec representation to a high-quality target distribution, and A2SB~\cite{kong2025a2sb} extended this to end-to-end 44.1\,kHz music restoration.
Diffusion vocoders such as DiffWave~\cite{kong2020diffwave} and Multi-Band Diffusion~\cite{sanroman2023multiband} generate waveforms conditioned on codec features, while flow matching approaches~\cite{le2023voicebox, welker2025flowdec} offer an alternative iterative framework.

% DONE
These methods do exhibit emergent coarse-to-fine behavior, where high noise levels correspond to coarse structure and low noise levels to fine detail, but this correspondence is implicit and not aligned to the codec's own discrete hierarchy.
The denoising schedule is independent of the RVQ layer decomposition.

\section{Methodology}
\label{sec:method}

\begin{figure*}[t]
	\centering
    \includegraphics[alt={Three-panel diagram of geometric iterative retrieval.
        Panel (a), training: a grid of discrete tokens for RVQ layers 1 to 8 enters
        the model, which predicts continuous latents for layers 2 to 9 in a single
        forward pass; predicted layer is supervised with a CLIP-style
        contrastive loss. Panel (b), inference: tokens of layers 1 to 4 enter the
        model, which predicts the layer-5 latent; nearest-neighbor lookup in the
        layer-5 codebook converts the latent to layer-5 tokens, which are appended
        to the input stack for the next step. Panel (c), architecture: at each
        sequence position, per-layer codebook embeddings are combined by a
        self-attention aggregator, processed by a DeBERTa-v3 encoder, and projected
        by an output matrix W out to the predicted latent of the next layer.},
    width=\textwidth]{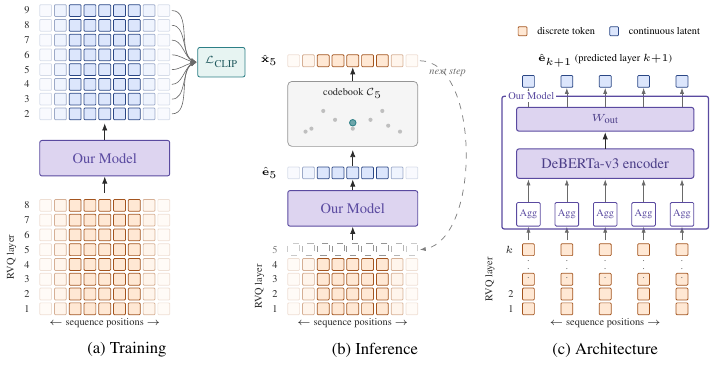}

	\caption{Combined view of geometric iterative retrieval.
		(a) During training, the model predicts layers $2, \dots, D$ from
		the masked input stack in a single forward pass, supervised by a
		CLIP-style contrastive loss per predicted layer.
		(b) At inference, layer $k+1$ is predicted from layers $1, \dots, k$,
		quantized by nearest-neighbor lookup against the layer-$k+1$ codebook,
		and concatenated for the next step.
		(c) Architecture: per-position codebook embeddings are aggregated by
		self-attention into one hidden state, processed by the DeBERTa-v3
		encoder, and projected by $W_{\text{out}}$ into the codebook space.
		Orange = discrete tokens, blue = continuous latents.}
	\label{fig:method_combined}
\end{figure*}

% DONE
We implement geometric iterative retrieval as a layer-wise prediction model over DAC's RVQ stack, trained with a contrastive retrieval objective and a self-attention aggregator (see Figure~\ref{fig:method_combined}c).
Figure~\ref{fig:method_combined}(a,b) illustrates the overall training and inference pipeline.

% Old two-panel figure replaced by fig:method_combined below.
% \begin{figure*}[t]
% 	\centering
% 	\begin{minipage}[b]{0.49\textwidth}
% 		\centering
% 		\resizebox{\textwidth}{!}{\input{assets/method.tex}}
% 		\caption{Training and Inference pipeline. During training, we predict layer $k+1$ from layers $1$ to $k$, for $k=1,\ldots,D-1$, computing the CLIP objective on the latent predictions. At inference, we only provide the first-layer tokens, and the model predicts one additional RVQ layer per step, autoregressively consuming its outputs until all $D$ layers are produced, after which the DAC decoder reconstructs the waveform.}
% 		\label{fig:method}
% 	\end{minipage}\hfill
% 	\begin{minipage}[b]{0.49\textwidth}
% 		\centering
% 		\resizebox{\textwidth}{!}{\input{assets/architecture}}
% 		\caption{Geometric iterative retrieval architecture. Each layer learns a separate embedding table. The first $k$ layer discrete tokens (orange) are aggregated across the RVQ dimension with a self-attention mechanism, then passed to a bidirectional DeBERTa-v3 encoder model. An output head projects them into $\mathbb{R}^d$ to predict the continuous latents (blue) of the $k+1$-th layer.}
% 		\label{fig:architecture}
% 	\end{minipage}
% \end{figure*}

\subsection{Problem Formulation}
\label{sec:problem}
% DONE
Given the first-layer codebook tokens $\mathbf{x}_1 \in \{1, \dots, V\}^T$ of an audio segment, the task is to predict the remaining codebook vectors $\mathbf{e}_2, \dots, \mathbf{e}_D \in \mathbb{R}^{T \times d}$ such that the decoded audio quality is maximized.
The prediction target is the sequence of per-layer codebook vectors $\mathbf{e}_k$ rather than discrete indices or the full pre-quantized embedding $\mathbf{z}$.
This choice is what enables our approach: operating in $\mathbb{R}^d$ grants access to the codebook's geometric structure, whilst the per-layer decomposition preserves a semantically meaningful iterative axis.

\subsection{Architecture}
\label{sec:architecture}
% DONE
The backbone of our model is a bidirectional DeBERTa-v3~\cite{he2023debertav3} transformer encoder with 12 layers and hidden dimension 1536.
Tokens at each layer are mapped to hidden states via a per-layer, learned embedding table.
A single prediction head projects the transformer output to the codebook dimension $d$.

\noindent\textbf{Self-Attention Codebook Aggregation.}
\label{sec:aggregator}
% DONE
Standard RVQ decoding reconstructs the latent as the sum $\hat{\mathbf{z}} = \sum_k \mathbf{e}_k$.
This additive assumption is an architectural constraint of the codec, not a property we need to preserve when conditioning on previously predicted layers.
We replace summation with an attention module on the embeddings of layers $1, \dots, k{-}1$.
This lets the model weight earlier layers non-uniformly and capture cross-layer dependencies that pure summation cannot represent.

\subsection{Training Objective}
\label{sec:training}

% DONE
All $D{-}1$ residual layers are predicted in a single forward pass with causal masking along the codebook axis, this results in one pass supplying ground-truth conditioning from lower layers while producing predictions for all higher layers simultaneously.

% DONE
The training loss is a CLIP-style symmetric contrastive loss~\cite{radford2021clip} between predicted and target codebook vectors.
For each layer $k$ and each position in the batch, the predicted vector $\hat{\mathbf{e}}_k$ must be closer in cosine similarity to the true codebook vector $\mathbf{e}_k$ than to any other codebook vector drawn from the batch.
A learnable temperature parameter controls the sharpness of the contrastive distribution.
%  This objective is the core mechanism that distinguishes our method from the paradigms surveyed in Section~\ref{sec:related_work}:

% \noindent\textbf{Geometry-aware gradient.} Unlike cross-entropy token prediction, the contrastive loss respects codebook geometry: a prediction near the target incurs less loss than one far away, providing gradient signal from near-misses that classification lacks.

% \noindent\textbf{Anchored to the codebook.} Unlike MSE or cosine regression to $\mathbf{z}$, the objective uses other codebook entries as explicit negatives. The model is pushed not merely toward the target but away from neighbors, which suppresses the mode-averaging that yields the over-smoothed predictions characteristic of one-step regression.

% \noindent\textbf{Discriminative, not generative.} Unlike diffusion, the objective retrieves the correct codebook vector rather than learning to sample from a full distribution over $\mathbf{z}$, which suffices for fidelity-driven resynthesis at a fraction of the computational cost.

\subsection{Inference}
\label{sec:inference}
% DONE
At inference, layers are predicted autoregressively along the codebook axis: $\hat{\mathbf{e}}_2$ is predicted from $\mathbf{x}_1$, then $\hat{\mathbf{e}}_3$ from $(\mathbf{x}_1, \hat{\mathbf{x}}_2)$, and so on through $\hat{\mathbf{e}}_D$,
where $\hat{\mathbf{x}}_k$ is the closest codebook in $\mathcal C_k$ to $\hat{\mathbf{e}}_k$.
The resulting stack $(\mathbf{x}_1, \hat{\mathbf{x}}_2, \dots, \hat{\mathbf{x}}_D)$ is passed through DAC's frozen per-layer output projections, summed to form the latent representation, and decoded by DAC's frozen decoder.
The full inference procedure requires $D{-}1$ deterministic steps, matching the step count of discrete coarse-to-fine prediction.

\section{Experimental Setup}
\label{sec:experimental_setup}
% DONE
We evaluate geometric iterative retrieval on codec restoration, the canonical
fidelity-driven resynthesis task: given the first RVQ layer of
a ground-truth codec token stream, predict the remaining layers, decode, and
measure the gap to the full $D$-layer reconstruction.

\subsection{Codec}
\label{sec:codec}
% DONE
All experiments use the 44.1\,kHz Descript Audio Codec~\cite{kumar2023dac} with
$D=9$ RVQ layers, codebook size $V=1024$, and codebook dimension $d_c=1024$. DAC performs lookup and quantization in a low-dimensional space, making the effective dimension we use for our predictions $d=8$. DAC's
encoder, quantizer, and decoder remain frozen throughout.

\subsection{Datasets}
\label{sec:datasets}
% DONE
Training and evaluation use three corpora spanning speech and music:
MTG-Jamendo~\cite{bogdanov2019mtgjamendo},
Common Voice~\cite{ardila2020commonvoice}, and FMA~\cite{defferrard2017fma}. Each corpus is tokenized with DAC once offline. We hold out a
10\,\% random validation split per corpus (seed 42). All benchmark numbers are
computed on these validation splits. Training mixes the three corpora with
sampling weights $10\!:\!1\!:\!2$ (Jamendo\,:\,CV\,:\,FMA) to balance
different average sample lengths. Each training sample is a 128-frame
random crop of the token sequence.

\subsection{Architecture and Training}
\label{sec:training_details}
% DONE
The backbone is a 12-layer DeBERTa-v3~\cite{he2023debertav3} encoder with hidden dimension 1536.
The codebook aggregator uses 4-head attention at the same dimension, and
the output head projects to the codebook dimension $d=8$. We train with
AdamW (learning rate $5\!\times\!10^{-5}$, weight decay $5\!\times\!10^{-2}$),
a linear warmup over the first 10k steps followed by cosine decay to
$10^{-6}$ over 200k total steps, batch size 16, and bf16 mixed precision.
The CLIP-style contrastive loss (Section~\ref{sec:training}) uses a learnable
temperature. All $D-1$ residual layers are predicted in one forward pass
via causal masking along the codebook axis.

\subsection{Baselines and Ablations}
\label{sec:baselines}
% DONE
We compare against various baselines and ablations. All models are trained
on the same datasets for 24 hours using four NVIDIA RTX A6000s.

\noindent\textbf{Naive first-layer decode.} Decode only the given layer through DAC. This is the
zero-cost lower bound for any resynthesis method.

\noindent\textbf{One-step continuous regression (OSR).}
A DeBERTa-v3 encoder of the same capacity as ours, trained to regress the
pre-quantized latent $\mathbf{z}$ in a single forward pass with
cosine-similarity loss. Instead of predicting a $d$-dimensional vector at each pass, it predicts a $(D-1)\times d$-dimensional vector that is chunked into $D-1$ codebook predictions.

\noindent\textbf{MSE one-step regression.} Same DeBERTa-v3 backbone as OSR
but trained with mean-squared-error loss directly on the summed residual
codebook latent $\sum_{k=2}^{D} \mathbf{e}_k$, and decoded at inference by
greedy per-layer nearest-neighbor assignment in the codebook.

\noindent\textbf{Token prediction (CE).} A 9-layer Llama-style decoder
transformer (hidden 1536, 24 attention heads)
trained with next-token cross-entropy over a vocabulary of $D \cdot V =
	9216$ offset tokens, where layer-$k$ indices are shifted by $k \cdot V$
before flattening. Layer $1$ is supplied as
context and the remaining indices are decoded layer-by-layer.

\noindent\textbf{Full-residual variant (ablation).} Same backbone and
contrastive objective as ours, but predicting the cumulative residuals
$\sum_{j = k+1}^D \mathbf{e}_j$ rather than $\mathbf{e}_{k+1}$.

\noindent\textbf{No-contrastive variant (ablation).} Same backbone and
per-layer target as ours, but trained with cosine-similarity regression to
the true codebook vector rather than a contrastive loss over the codebook.

\noindent\textbf{Additive aggregator (ablation).} Same backbone as ours, but
adds the codebook embeddings together instead of using the self-attention
mechanism.

\subsection{Evaluation Protocol}
\label{sec:eval_protocol}
% DONE
We evaluate every method
on the validation split of each of the three corpora. We sample 500 clips uniformly at random per
corpus. Inputs are the
ground-truth first-layer codes, and the reference is the 9-layer DAC
reconstruction of the same clip. The gap therefore reflects only
resynthesis error, not DAC's own quantization loss.
We additionally report a
layer-progression analysis in Section~\ref{sec:results_layer_progression}.

% TO BE IMPROVED
We use three objective metrics for evaluating predictions, log-spectral distance (LSD)~\cite{gray1976lsd}, SI-SDR~\cite{leroux2019sisdr}, and Fr\'{e}chet Audio Distance (FAD)~\cite{kilgour2019fad,hershey2017cnn}. We use LSD as our primary objective metric.

% \subsection{Metrics}
% \label{sec:metrics}
% We report three metrics, each capturing a complementary failure mode.

% \textbf{Log-spectral distance (LSD)} is our primary metric. For STFT magnitudes
% $S_\text{pred}, S_\text{ref}$ with $n_\text{fft}=2048$ and hop 512, we compute
% \[
%     \text{LSD} \;=\; \frac{1}{T}\sum_t \sqrt{\frac{1}{F}\sum_f
%         \left(20 \log_{10}\frac{|S_\text{pred}[t,f]|}{|S_\text{ref}[t,f]|}\right)^2}
% \]
% (in dB; lower is better). LSD is phase-invariant and penalizes exactly the
% spectral smoothing failure mode of one-step regression.

% \textbf{Scale-invariant signal-to-distortion ratio (SI-SDR)}~\cite{leroux2019sisdr}
% (higher is better, in dB) is phase-sensitive and complements LSD by penalizing
% methods that match the spectrum but misalign in time.

% \textbf{Fr\'echet Audio Distance (FAD)}~\cite{kilgour2019fad} (lower is better) is a
% set-level distributional metric computed from VGGish embeddings. It captures
% perceptual plausibility: the predicted set is scored against the full-layer
% set of the same clips, so a method that produces artefact-free but
% mode-collapsed outputs can win on SI-SDR and LSD while losing on FAD.

% Per-pair metrics (LSD, SI-SDR) are reported as means across all 1500 val
% samples; FAD is a single number per $(\text{method}, n_\text{given})$ cell.

\section{Results}
\label{sec:results}
\subsection{Codec Restoration}
\label{sec:results_restoration}
% DONE
Table~\ref{tab:main_results} reports the three metrics for all baselines and our method.
The naive first-layer decode establishes the
zero-cost floor: the gap it leaves to the full-layer reference quantifies
how much information is carried by the residual layers and therefore how
much a resynthesis method can possibly recover.

% DONE
\begin{table}[t]
	\centering
	\small
	\begin{tabularx}{\linewidth}{l Y Y Y}
		\toprule
		\textbf{Method} & LSD\,$\downarrow$           & SI-SDR\,$\uparrow$          & FAD\,$\downarrow$ \\
		\midrule
		Naive           & $\underline{10.82\!\pm\!0.08}$ & $-1.88\!\pm\!0.24$             & 0.99                  \\
		CE              & $12.42\!\pm\!0.20$             & $-6.93\!\pm\!0.30$             & 2.46                  \\
		OSR             & $11.03\!\pm\!0.07$             & $\mathbf{-0.22\!\pm\!0.27}$    & \textbf{0.61}         \\
		MSE             & $12.86\!\pm\!0.22$             & $-7.83\!\pm\!0.30$             & 3.08                  \\
		\textbf{Ours}   & $\mathbf{10.61\!\pm\!0.07}$    & $\underline{-0.63\!\pm\!0.28}$ & \underline{0.94}      \\
		\bottomrule
	\end{tabularx}
	\caption{Codec restoration results on the combined
		val split (1500 clips). Lower is better for LSD and FAD, higher is better for SI-SDR.
		Best per column in \textbf{bold}, second-best \underline{underlined}.
		LSD and SI-SDR are reported as mean $\pm$ half-width of the 95\,\% normal CI over the
		1500 clips. FAD is a single set-level number per cell and admits no per-sample CI.
		\emph{OSR} is one-step, layer-wise regression under cosine-similarity loss.
		\emph{MSE} is the same backbone
		trained with MSE against the residual codebook latent.}
	\label{tab:main_results}
\end{table}

% DONE
Three observations are noteworthy. (i)~\textbf{Our method wins the primary metric against every learned baseline.} It attains the best LSD and
beats the CE token-prediction baseline and the MSE one-step baseline on all
three metrics. OSR achieves the best SI-SDR and FAD on this table. Our
advantage on the objective metrics is concentrated on LSD, and the listening
study (Section~\ref{sec:results_subjective}) is what most clearly separates
the two methods.
(ii)~\textbf{The continuous single-step cell is split by the choice of
	loss.} OSR (cosine-similarity) sits near the naive first-layer floor on LSD
and beats it on SI-SDR/FAD, whereas MSE performs poorly. This is the conditional-mean
minimizer of Section~\ref{sec:regression} made concrete. Under a
multimodal posterior, MSE regresses toward zero rather than committing to
any mode. Due to the highly non-linear nature of the output projection,
this leads to poor reconstruction.
(iii)~\textbf{Discrete iterative prediction underperforms
	continuous iterative prediction by a wide margin.} The CE baseline underperforms with $1.81$\,dB compared to our approach on LSD and $6.30$\,dB worse on SI-SDR: a
geometry-blind classification loss cannot exploit near-misses, confirming
the argument of Section~\ref{sec:token_prediction}.

\textbf{Against the naive first-layer decode} our approach wins on all three
metrics: LSD ($10.61$ vs.\ $10.82$), SI-SDR ($-0.63$ vs.\ $-1.88$), and FAD
($0.94$ vs.\ $0.99$). The
layer-progression analysis of Section~\ref{sec:results_layer_progression}
localises this gain to the first two predicted layers. Beyond $K{=}3$,
accumulated prediction error outweighs the residual signal each new layer
is meant to encode.

\subsection{Layer-Progression Analysis}
\label{sec:results_layer_progression}

To isolate the contribution of each predicted residual layer, we decode our
predictions truncated to the first $K$ layers for $K \in \{1, \dots, D\}$ and
score each truncation against the full $D$-layer reference (Table~\ref{tab:layer_progression}). This exposes how
much of the overall gain comes from each step of the iterative retrieval
process.

\newcommand{\dpct}[1]{{\scriptsize\textcolor{black!55}{(#1)}}}
\begin{table}[t]
	\centering
	\small
	\setlength{\tabcolsep}{4pt}
	\begin{tabular*}{\linewidth}{@{\extracolsep{\fill}}l c c c@{}}
		\toprule
		$K$            & LSD\,$\downarrow$                       & SI-SDR\,$\uparrow$                        & FAD\,$\downarrow$                           \\
		\midrule
		1 (input only) & 10.81                                   & $-1.88$                                   & 0.99                                        \\
		2              & 10.26 \dpct{$-5.1\%$}                   & $-0.34$ \dpct{$+82\%$}                    & \textbf{0.43} \dpct{$-57\%$}                \\
		\textbf{3}     & $\mathbf{10.20}$ \dpct{$\mathbf{-5.6\%}$} & $\mathbf{-0.22}$ \dpct{$\mathbf{+88\%}$} & $\mathbf{0.43}$ \dpct{$\mathbf{-57\%}$}    \\
		4              & 10.23 \dpct{$-5.4\%$}                   & $-0.30$ \dpct{$+84\%$}                    & 0.56 \dpct{$-43\%$}                         \\
		5              & 10.34 \dpct{$-4.3\%$}                   & $-0.41$ \dpct{$+78\%$}                    & 0.69 \dpct{$-30\%$}                         \\
		6              & 10.41 \dpct{$-3.7\%$}                   & $-0.50$ \dpct{$+73\%$}                    & 0.77 \dpct{$-22\%$}                         \\
		7              & 10.49 \dpct{$-3.0\%$}                   & $-0.56$ \dpct{$+70\%$}                    & 0.81 \dpct{$-18\%$}                         \\
		8              & 10.55 \dpct{$-2.4\%$}                   & $-0.63$ \dpct{$+66\%$}                    & 0.88 \dpct{$-11\%$}                         \\
		9              & 10.61 \dpct{$-1.9\%$}                   & $-0.64$ \dpct{$+66\%$}                    & 0.94 \dpct{$-5\%$}                          \\
		\bottomrule
	\end{tabular*}
	\caption{Reconstruction quality as a function of the number of
		predicted residual layers $K$ (layer $1$ is the ground-truth input,
		layers $2, \dots, K$ are our predictions). All three objective metrics
		improve from $K{=}1$ through $K{=}3$ and then degrade monotonically,
		because each additional residual layer brings progressively less
		spectral signal while still injecting prediction noise of roughly the
		same magnitude. Beyond $K{=}3$, the noise outweighs the signal.
		Parenthesized values report the percentage change relative to the
		$K{=}1$ input-only baseline. Best per column in \textbf{bold}.}
	\label{tab:layer_progression}
\end{table}

% DONE
\textbf{All three metrics peak at $K{=}3$ or earlier}, and degrade monotonically thereafter, with
FAD more than doubling from $K{=}3$ ($0.43$) to $K{=}9$ ($0.94$).
Each predicted layer
contributes both signal (the next codec residual) and noise from its own
prediction error, and past $K{=}3$ the second outweighs the first. The truncated
$K{=}3$ decode therefore dominates the full $K{=}D$ decode on every spectral metric.
One possible explanation is that, since codebooks at each layer have uniform length~\cite{kumar2023dac},
the model can only choose the direction, not the norm, of the next prediction.
This forces the model to commit to a mode, increasing spectral distance on average.
We investigate whether this truncation also matches perceptually in the  next section.

\subsection{Subjective Evaluation}
\label{sec:results_subjective}
% DONE
We complement the objective metrics with a double-blind human evaluation on a
diverse selection of nine 10-second clips drawn from the validation splits: 2 from MTG-Jamendo, 3 from FMA and 4 from Common Voice. Participants were recruited online\footnote{https://www.mabyduck.com/}. For each clip, listeners rated the six conditions of Table~\ref{tab:human_eval} on a 0--100 quality
scale against the original audio as an explicit reference. Nineteen listeners
each rated all nine clips, yielding $19 \times 9 = 171$ within-listener-clip
paired trials per condition. We additionally include a truncated variant of our method (K{=}3) motivated by the layer progression of Section~\ref{sec:results_layer_progression}, which decodes only the first two predicted residual layers.
% DONE
Table~\ref{tab:human_eval} reports the per-condition mean and 95\,\% confidence
interval, together with the paired comparison against our method.
\begin{table}[t]
	\centering
	\small
	\begin{tabularx}{\linewidth}{p{1.6cm} P{2.2cm} Y Y}
		\toprule
		\textbf{Condition} & \textbf{Mean (95\% CI)}    & $\Delta$ & \textbf{W/L}         \\
		\midrule
		Naive              & 23.8 [21.0, 26.7]          & $+31.4$  & 19\,/\,148           \\
		CE                 & \phantom{0}9.1 [7.5, 10.7] & $+45.5$  & \phantom{0}6\,/\,164 \\
		MSE                & \phantom{0}7.6 [5.7, 9.6]  & $+46.8$  & \phantom{0}7\,/\,164 \\
		OSR                & 43.4 [39.7, 47.1]          & $+11.2$  & 48\,/\,117           \\
		Ours (K=3)         & 49.3 [45.7, 52.8]          & $+5.3$   & 64\,/\,\phantom{0}99 \\
		\textbf{Ours}      & \textbf{54.0 [50.1, 57.9]} & ---      & ---                  \\
		\bottomrule
	\end{tabularx}
	\caption{Double-blind human evaluation (19 raters $\times$ 9 clips).
	Higher is better. $\Delta$ is the mean per-trial difference (Ours $-$
	baseline). \textbf{W}ins/\textbf{L}osses count the trials in which the
	baseline scored above or below Ours (ties omitted). All five pairwise
	differences favor our method and are significant under a Wilcoxon signed-rank
	test (Naive, CE, MSE with $|z| > 9.7$, $p < 10^{-21}$; OSR with $|z| = 5.5$, $p = 3.4{\times}10^{-8}$; Ours (K{=}3) with $|z| = 3.9$, $p = 1.1{\times}10^{-4}$).}
	\label{tab:human_eval}
\end{table}
% TO BE IMPROVED
The subjective ranking underscores the findings of the objective evaluations.
Our method is preferred over every baseline by a $\sim\!11$--$47$ point margin
in mean rating, with raters preferring it to OSR on 117 of 165
non-tied trials and to the discrete CE baseline on 164 of 170. The two
cells of Table~\ref{tab:paradigm_grid} that we argue to be
fundamentally limited (discrete CE and continuous single-step regression
under MSE) are also the lowest-rated by listeners, scoring \emph{below} the
naive first-layer decode despite being trained predictors. This is a
clear instantiation of the failure modes identified in
Section~\ref{sec:related_work}. OSR avoids
both pitfalls, but still trails our method by $11.2$ points on average,
isolating the contribution of the iterative
codec-native retrieval paradigm itself. The truncated K{=}3 variant, which Section~\ref{sec:results_layer_progression} flagged as a
natural next step on the basis of objective metrics, trails the full K{=}9 decode by $5.3$ points
($p \approx 1.1{\times}10^{-4}$). This reverses the direction of the objective
layer-progression. Spectral metrics prefer $K{=}3$, listeners prefer $K{=}D$. Empirically,
$K=D$ produces a smoother sound, with less harsh and crackly artifacts.
We read this as a feature of the codec-native iterative paradigm rather than
a defect, in that lower $K$ is better for overall spectral fidelity (pure retrieval), whilst
using the entire $D$ layers yields better sounding samples, which is an advantage for generative tasks.

\subsection{Ablations}
\label{sec:results_ablations}
% DONE
We evaluate the three ablations from Section~\ref{sec:baselines},
each isolating one component of the proposed method.
The results can be seen in Table~\ref{tab:ablations}.
% DONE
\begin{table}[t]
	\centering
	\small
	\setlength{\tabcolsep}{4pt}
	\begin{tabular*}{\linewidth}{@{\extracolsep{\fill}}l c c c@{}}
		\toprule
		\textbf{Variant} & LSD\,$\downarrow$         & SI-SDR\,$\uparrow$        & FAD\,$\downarrow$ \\
		\midrule
		\textbf{Ours}    & $\mathbf{10.61 \pm 0.07}$    & $\mathbf{-0.63 \pm 0.28}$    & \textbf{0.94}    \\
        \midrule
		Full-residual    & $12.29 \pm 0.22$             & $-4.22 \pm 0.32$             & 1.14             \\
		No-contrastive   & $\underline{10.75 \pm 0.07}$ & $\underline{-1.17 \pm 0.31}$ & 0.99             \\
		Additive         & $11.74 \pm 0.09$             & $-1.24 \pm 0.30$             & \underline{0.96} \\
		\bottomrule
	\end{tabular*}
	\caption{Ablation results on the combined validation
		split. Each variant perturbs one component, the prediction target
		(per-layer $\mathbf{e}_k$ vs.\ cumulative residual
		$\sum_{j = k+1}^D \mathbf{e}_j$), the loss (contrastive retrieval vs.\
		cosine regression to the codebook vector), or the layer-aggregation
		mechanism (self-attention vs.\ additive summation). Best per column in
		\textbf{bold}, second-best \underline{underlined}. LSD and SI-SDR are
		reported as mean $\pm$ half-width of the 95\,\% normal CI over the
		1500 clips. FAD is a single set-level number per cell.}
	\label{tab:ablations}
\end{table}

\noindent\textbf{Full-Residual vs.\ Per-Layer Target.}
\label{sec:results_ablation_target}
% DONE
Predicting cumulative residuals rather than per-layer codebook vectors is clearly worse, with a $1.7$\,dB LSD penalty and a $3.6$\,dB SI-SDR penalty. The per-layer target is what allows the contrastive retrieval objective to be anchored to individual codebook entries. Under the cumulative-residual target, the set of negatives (all codebook vectors in $\mathcal{C}_k$) is no longer geometrically aligned with the target (a sum of codebook vectors), and the retrieval signal collapses. This confirms that the per-layer decomposition is
load-bearing, not an incidental architectural choice.

\noindent\textbf{Contrastive vs.\ Cosine-Regression Objective.}
\label{sec:results_ablation_loss}
% DONE
Replacing contrastive retrieval loss with direct cosine regression
against the true codebook vector yields a small but consistent penalty:
$0.14$\,dB on LSD, $0.54$\,dB on SI-SDR, and $0.05$ on FAD
(Table~\ref{tab:ablations}). The gap is smaller than the full-residual
ablation because the per-layer target already constrains the solution to
the codebook geometry. The contrastive loss sharpens this by pushing
predictions away from other codebook entries, not just toward the target.

\noindent\textbf{Additive vs.\ Self-Attention Aggregator.}
\label{sec:results_ablation_aggregator}
% DONE
Replacing the self-attention aggregator with the additive combination of
previously predicted layer embeddings, as in~\cite{borsos2023soundstorm},
costs $1.13$\,dB on LSD, $0.61$\,dB on SI-SDR, and $0.02$ on FAD
(Table~\ref{tab:ablations}). Keeping
the per-layer target and contrastive loss intact preserves most of the
quality, but the additive prior is still noticeably worse than the
attention-based combination at higher RVQ layers, where deviations from
exact summation accumulate. This is consistent with the motivation of
Section~\ref{sec:aggregator}, in that the additive assumption is an architectural
constraint of the codec's decoder, not a property the model has to inherit
when conditioning on its own previous predictions.

\section{Conclusion}
\label{sec:conclusion}
% DONE
We argued that the design space for codec resynthesis is a
two-dimensional grid spanning prediction space and refinement strategy, and
that the continuous-iterative cell, occupied to date only by diffusion,
admits a second instantiation that follows the codec's own RVQ
hierarchy rather than an externally imposed noise schedule. We
instantiated this paradigm as \emph{geometric iterative retrieval},
per-layer prediction in continuous codebook space trained with a CLIP-style
contrastive objective and conditioned on previous layers through a
learned non-additive self-attention aggregator. On DAC codec
restoration, our method attains the best LSD against every learned
baseline and is preferred by listeners, with the
CE and MSE baselines rated below the naive first-layer decode.
Ablations confirm that the per-layer target, the contrastive objective,
and the self-attention aggregator each contribute.
% \noindent\textbf{Limitations.} Our evaluation has three principal
% limitations. First, all experiments use a single codec (DAC); our claim
% that geometric iterative retrieval is codec-agnostic in principle is not
% empirically tested on EnCodec, SoundStream, or non-RVQ codecs. Second, we
% report no head-to-head comparison with diffusion or flow-matching codec
% resynthesis: the closest published
% systems~\cite{liu2024codec_resyn,kong2025a2sb} target different codecs and
% are not publicly released, and a faithful reproduction is beyond the
% scope of this work. Our results therefore position geometric iterative
% retrieval against the discrete-iterative and continuous-single-step
% paradigms, not against the continuous-iterative diffusion cell, which we
% leave to future work. Third, on the objective metrics our advantage over
% the OSR cosine-regression baseline is concentrated on LSD; OSR is
% competitive on SI-SDR and FAD, and the clearest separation between the
% two methods comes from the listening study, which covers 19 raters and 9
% clips. The OSR vs.\ Ours and $K{=}3$ vs.\ $K{=}D$ comparisons in
% particular would benefit from a larger panel.
% DONE
The layer-progression analysis exposes a tension between objective and
subjective evaluation. All three objective metrics peak at $K{=}3$ and
then degrade monotonically, yet listeners prefer the full $K{=}9$
decode to the truncated variant. Higher residual layers contribute
audible quality that LSD, SI-SDR, and FAD do not capture, which points
to a need for codec-resynthesis-aware objective metrics that go beyond
spectral fidelity. Beyond DAC, geometric iterative retrieval is
codec-agnostic, and extending it to other RVQ codecs and
to settings where the coarse layers are themselves generated rather
than given is a natural next step.

% For BibTeX users:
\bibliography{references}

\end{document}